\documentclass[cameraready]{Interspeech}

\title{MagpieTTS-LF: Inference-Time Long-Form Speech Generation Without Training on Long-Form data}

\author[affiliation={1}]{Subhankar}{Ghosh}
\author[affiliation={1}]{Jason}{Li}
\author{Paarth Neekhara$^{1}$, Shehzeen Hussain$^{1}$, Ryan Langman$^{1}$, Xuesong Yang$^{1}$,  Roy Fejgin$^{1}$}

\address{
    $^1$ NVIDIA Corporation, USA
}

\email{\{subhankarg,jasoli,pneekhara,shehzeenh,rlangman,xueyang,rfejgin\}@nvidia.com\vspace{-0.7cm}}

\keywords{Text-to-Speech, Speech Synthesis, Speech LLM, Long-form Generation}

\usepackage{comment}
\usepackage{amssymb}
\usepackage{float}

\begin{document}

\maketitle

\begin{abstract}
    Neural Text-to-Speech (TTS) systems achieve remarkable quality on short utterances but long-form speech generation shows prosodic drift, speaker inconsistencies and sentence boundary artifacts. Existing approaches either compress sequences, increase context length or naively concatenate independently synthesized chunks. We present an inference-time approach called \textbf{MagpieTTS-LF} that enables MagpieTTS to produce coherent long-form speech without model retraining. Our method introduces three key innovations: (1) soft attention priors to guide monotonic alignment while preserving past and future context; (2) a stateful inference algorithm that maintains context across sentence chunks, ensuring prosodic continuity; (3) history-aware text encoding that uses past text for discourse-level prosodic planning. Experiments on long texts show significant improvements in long-range intelligibility, prosodic coherence, speaker consistency, and boundary naturalness compared to other baselines.
\end{abstract}


\section{Introduction}

While advancements in large-scale generative modeling in TTS has enabled unprecedented naturalness and speaker similarity, yet most of the methods suffer from hallucinations, prosodic drift, and boundary artifacts as generation length grows. State-of-the-art models like Tortoise TTS~\cite{betker2023tortoise}, VALL-E 2~\cite{chen2024valle2}, VALL-E R~\cite{valle_r}, NaturalSpeech 2/3~\cite{shen2023naturalspeech, ju2024naturalspeech}, VoiceBox~\cite{NEURIPS2023_2d8911db}, XTTS~\cite{casanova2024xtts}, Qwen3-TTS~\cite{hu2026qwen3tts}, MagpieTTS~\cite{neekhara24_interspeech, koel2025} and CosyVoice~\cite{du2024cosyvoice} produce extremely natural speech on 2-20 second utterances but offer no native mechanism for paragraph length speech generation. When generating longer length speech, these systems default to sentence-level chunking followed by concatenation. This strategy that produces characteristic artifacts including energy discontinuities and warbles at boundaries, inconsistent speaking rate across segments, and loss of prosodic patterns such as intonation shift.

The literature shows three paradigms for extending generation length, each with distinct limitations. Sequence compression approaches reduce the number of tokens per second of audio to fit more speech within fixed context windows. VibeVoice \cite{peng2025vibevoice} achieves a compression at 7.5 Hz, enabling up to 90 minutes of speech generation in a single pass, however it sacrifices temporal resolution, representing each $\sim133 ms$ of audio with a single token. SpeechSSM \cite{miyazaki2022speechssm} uses state-space models for theoretically infinite extrapolation. Streaming and block-wise methods such as CosyVoice 2 \cite{du2024cosyvoice2} employ block-wise attention masks to enable chunk-wise generation with bounded memory. However, these binary masks create hard information cutoffs rather than graceful gradients. Moreover, all streaming approaches require training time architectural modifications; the chunking strategy is baked into the model. Cross-sentence prosody models like HiGNN-TTS \cite{guo2023hignn}, and Context-Aware Memory \cite{li2025contextawarememory} demonstrate that text context from neighboring sentences improves prosodic naturalness. Yet these approaches require dedicated graph networks, or memory modules trained jointly with the TTS system, making these improvements difficult to apply on existing deployments.

We present long-form speech generation algorithm \textbf{MagpieTTS-LF}\footnote{Open-source code https://github.com/NVIDIA-NeMo/NeMo}, addressing four converging gaps in the literature. First, our approach operates entirely at inference time, requiring no architectural changes or retraining existing MagpieTTS model. Second, we introduce soft attention priors that maintain non-zero weights on past and future tokens, preserving a gradient of attention, which does not completely suppress distant context. Our soft priors allow the model to retain long-range information while still focusing computation on locally relevant positions. Third, our stateful inference algorithm maintains attention prior states and encoder context across independently generated chunks, creating continuity between segments. Fourth, we leverage past text history by passing them through text encoder for prosodic planning, utilizing the model's native text representations rather than requiring specialized context modules.

Our key contributions can be summarized as follows:

\begin{itemize}  
\vspace{-0.05cm} 
\item We propose an inference-time soft attention prior mechanism that guides the auto-regressive generation toward monotonic alignment while preserving useful context from distant tokens. Unlike binary masking approaches, our priors maintain non-zero weights on past and future positions, enabling graceful information decay rather than hard cutoffs.

\item We introduce a stateful chunk generation algorithm that carries attention prior states, encoder hidden states, and text history across sentence boundaries. This enables coherent prosody in long-form synthesis without the memory overhead of single-pass generation or the discontinuities of naive chunk concatenation.

\item We demonstrate that inference-time soft priors combined with cross-chunk state propagation achieve significant improvements in long-range prosodic coherence, speaker consistency over multi-minute durations, and boundary naturalness.

\item 
We present a structured comparison against state-of-the-art models, spanning neural codec language models, large-scale multimodal architectures, and streaming diffusion frameworks.

\item We introduce \textbf{Long-Form HifiTTS dataset}, a benchmark dataset for evaluating long-form speech synthesis, designed to measure prosodic continuity, speaker persistence, and boundary robustness under extended generation settings.
\end{itemize}

\begin{figure}
    \centering
    \includegraphics[width=0.9\columnwidth]{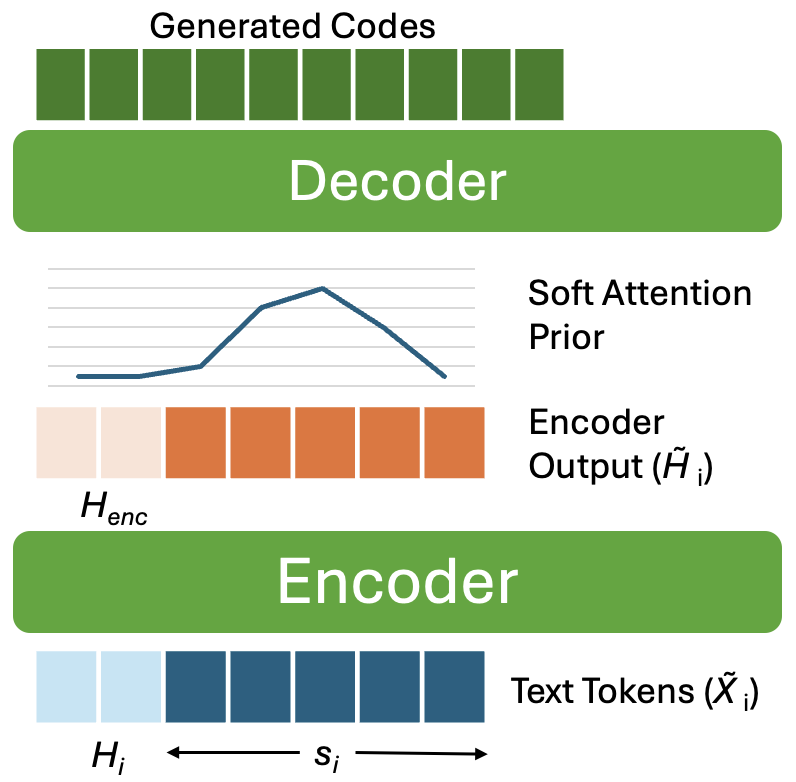}
    \caption{Stateful chunk generation for sentence $s_i$. History tokens $H_i$ are prepended to $s_i$ to form encoder input $\tilde{X}_i$ and the encoder output is concatenated with cached states $H_{\text{enc}}$ to produce $\tilde{H}_i$. A soft attention prior encourages monotonic alignment during decoding while preserving long-range context across chunk boundaries.}
    \label{fig:longformimage}
\end{figure}

\section{Methodology}

In this section, we present an inference-time approach for long-form speech generation that enables any chunk-based encoder-decoder TTS system to produce coherent speech without model retraining. In section \textbf{\ref{arch}}, we briefly go over the MagpieTTS model architecture. In section \textbf{\ref{softattention}}, we describe the soft attention priors that guide generation toward monotonic alignment while preserving context from distant tokens and section \textbf{\ref{chunk}} talks about a stateful chunk generation algorithm that maintains attention states and encoder context across sentence boundaries that is at the core of this approach.

\subsection{MagpieTTS Architecture} \label{arch}
The MagpieTTS model follows Koel-TTS~\cite{koel2025}, which is an encoder-decoder transformer architecture that operates on discrete audio tokens produced by a neural audio codec~\cite{casanova2025nanocodec}. The encoder processes text through a stack of self-attention layers, producing contextual representations that guide audio generation. The decoder auto-regressively generates discrete audio tokens by attending to both the encoded text and any provided audio context for voice cloning. During training, MagpieTTS employs Connectionist Temporal Classification (CTC) loss and learned attention priors to enforce monotonic cross-attention between text and audio, preventing hallucinations such as word repetition, skipping, or misalignment.

The standard model operates on single utterances; for longer inputs exceeding the generation time of ~20 seconds, the text must be split into chunks and processed independently. This causes the cross-sentence coherence to be lost and boundary artifacts.

\subsection{Inference-Time Soft Attention Priors} \label{softattention}

We introduce soft attention priors that maintain non-zero weights across the full context while focusing on the locally relevant positions, this does not suppress distant positions completely unlike binary attention masks. At each decoding step $t$, we compute (1) the text position $T_t$ that has the highest cross-attention score at decoding timestep $t-1$, (2) a prior distribution $P_t \in \mathbb{R}^N$ over encoder positions based on the expected monotonic alignment:

$$    P_t[i] = w_j , \quad \text{if } i \in \{T_t-1, T_t, T_t+1, T_t+2, T_t+3 \}$$
$$P_t[i] = eps, \quad \text{if } i \notin \{T_t-1, T_t, T_t+1, T_t+2, T_t+3 \}$$

Here i is the index of the text tokens, $w = (w_{-1}, w_{0}, w_{1}, w_{2}, w_{3})$ are the experimentally determined fixed weight vector that defines the attention of the current segment of the text, and $eps$ is the epsilon attention value for distant positions. This ensures that distant encoder positions still receive non-zero weight, preserving gradual context rather than hard cutoffs. The modified attention $\tilde{A}_t$ becomes:

\[\tilde{A}_t = \mathrm{softmax}\left(\frac{Q_t K^\top}{\sqrt{d}} + \lambda \log P_t\right)\]

where $\lambda$ controls the prior strength and is experimentally determined. This formulation allows the model to leverage its learned attention patterns while being gently guided toward monotonic alignment. Unlike the binary masks, our soft priors preserve information from distant tokens—useful for long-range dependencies

\subsection{Stateful Chunk Generation Algorithm} \label{chunk}

To generate long-form speech, we process long text in sentence-level chunks while maintaining information across chunk boundaries. The state comprises three components:

\begin{itemize}
    \item \textbf{History Text Tokens} $H_{text} \in \mathbb{R}^{B\times K}$: The final $K$ text tokens from the previous chunk, prepended to the current chunk's input to provide linguistic context for prosodic planning
    \item \textbf{History Encoder Context} $H_{enc} \in \mathbb{R}^{B\times K\times d}$: The corresponding encoder hidden states for the history tokens, concatenated with the current chunk's encoder output to provide continuous representations across boundaries. The history text tokens are first appended, then the encoder outputs at the corresponding positions are discarded, and finally history encoder context is appended
    \item \textbf{Attention Tracking}: A record of the last attended text positions from the previous chunk, used to initialize the soft attention prior in the current chunk, ensuring smooth prosodic transitions. $\tau$ is the initial soft attention prior.
\end{itemize}

The generation algorithm, also shown in Figure \ref{fig:longformimage}, works as follows. Given a long input text, we use punctuation-aware splitting to split into sentences $S = \{s_0, s_1, \ldots, s_M\}$. We initialize an empty state and process each sentence iteratively:

\begin{itemize}
    \item \textbf{Prepare Context}: Concatenate history text tokens with current sentence tokens: $\tilde{X}_i = [H_{\text{text}} ; s_i]$. Encode and concatenate with history encoder context: $\tilde{H}_i = \left[ H_{\text{enc}} ; \mathrm{Encoder}(s_i) \right]$.
    \item \textbf{Prior-Guided Generation}: Generate audio tokens auto-regressively using soft attention priors initialized from $\tau$. The prior guides attention to start where the previous chunk ended, ensuring prosodic continuity. At every auto-regressive step the prior is updated to guide the model to maintain monotonicity.
    \item \textbf{Update and Maintain State}: Save the final $K$ tokens and their encoder states as new history. Record final attention prior weights as new $\tau$.
    \item \textbf{Save Generated Code}: Once end of speech token has been detected for the given sentence, save the generated codes in the current iteration. Once all the sentences have been generated, these codes would be concatenated to get the entire audio code sequence for the total input text.
\end{itemize}

\subsection{Long-Form HifiTTS dataset}

We construct a long-form evaluation benchmark concatenating paragraphs from Multilingual LibriSpeech (MLS)~\cite{Pratap2020MLSAL} to form 20 passages of approximately 3-4 minutes each. We estimate the duration by assuming 135 words per minute rate of speech in English. We normalize and clean the text resolving acronyms, numerical values.

\section{Experiments and Results}

We evaluate MagpieTTS-LF\footnote{Website: \href{https://magpietts-lf.github.io/}{https://magpietts-lf.github.io/}} long-form generation against the state-of-the-art baselines across three dimensions: alignment robustness over long sequences, prosodic continuity at chunk boundaries, and speaker identity consistency, naturalness over long sequences. Together, these metrics capture the major challenges of long-form synthesis identified in our literature review.

\subsection{Evaluation Setup}
We evaluate on a curated dataset comprising of 20 long texts in English. We compare against Qwen3-TTS~\cite{hu2026qwen3tts}, VibeVoice-TTS~\cite{peng2025vibevoice}, X-TTS~\cite{casanova2024xtts} inference. All inference are run on single A6000 GPU. We use Whisper-Large~\cite{radford2023whisper} for ASR. For Long-form speech generation algorithm with MagpieTTS-LF we use the 0.1 as the value of $eps$, $w = (0.2, 0.8, 1.0, 0.8, 0.2)$, temperature of $0.7$, $\lambda = 1.0$ and cfg scale of 2.5. For VibeVoice and Qwen3-TTS, we synthesize speech either by segmenting text into sentences using punctuation-aware segmentation and concatenating the resulting audio, or by processing the text unsegmented up to the model's maximum input length, and report whichever configuration performs better.

\subsection{Intelligibility and Speaker Similarity}

To measure alignment robustness over time, we compute Word Error Rate (WER) and Character Error Rate (CER). We generate transcriptions of the generated audio using Whisper-Large. We also compute speaker similarity (SSIM) by extracting speaker embeddings from reference speaker audio and generated speech using Titanet \cite{koluguri2022titanet} and WavLM \cite{Chen_2022}, and then calculating cosine similarity.

\textbf{Results:} 
As shown in Table \ref{tab:intelligibility}, our proposed method in MagpieTTS-LF achieves significantly lower WER and CER compared to all other systems. While XTTS and Qwen3-TTS demonstrate moderate degradation, most likely due to error accumulated at chunk boundaries and lack of past context, VibeVoice shows significant degradations. This might suggest that extreme token compression might lead to low intelligibility over extended sequences. Through these results, we can see that our inference-time stateful approach maintains robustness across long sequences while models relying simply on naive chunking and aggressive compression compound errors as sequence length increases.

We also find that SSIM (WavLM) is the highest for MagpieTTS-LF, Qwen3-TTS is a close second, and XTTS suffers from the naive chunking and concatenation leads to inconsistent speaker characteristics across boundaries. VibeVoice suffer from low SSIM showing that token compression might lead to loss of speaker characteristics. Although Qwen3-TTS achieves the highest SSIM (TitaNet) score, its margin over MagpieTTS-LF is not statistically significant, so we do not discuss this difference further.

\begin{table}[t]
    \centering
    \caption{Intelligibility evaluation on the long-form HiFiTTS 1-hour subset. Word Error Rate (WER) and Character Error Rate (CER) are computed using Whisper-Large. Lower values indicate higher intelligibility.}
    \label{tab:intelligibility}
    \resizebox{\columnwidth}{!}{
    \begin{tabular}{lcccc}
        \toprule
        Model & WER $\downarrow$ & CER $\downarrow$ & SSIM (TitaNet) $\uparrow$ & SSIM (WavLM) $\uparrow$ \\
        \midrule
        \textbf{MagpieTTS-LF}  & \textbf{0.025} & \textbf{0.012} & 0.79 $\pm$ 0.02 & \textbf{0.979 $\pm$ 0.002} \\
        XTTS~\cite{casanova2024xtts}       & 0.051 & 0.035 & 0.69 $\pm$ 0.06 & 0.929 $\pm$ 0.042 \\
        Qwen3-TTS~\cite{hu2026qwen3tts}  & 0.045 & 0.028 & \textbf{0.80 $\pm$ 0.09} & 0.958 $\pm$ 0.025 \\
        VibeVoice\cite{peng2025vibevoice}  & 0.115 & 0.105 & 0.53 $\pm$ 0.15 & 0.848 $\pm$ 0.162 \\
        \bottomrule
    \end{tabular}
    }
\end{table}

\subsection{Prosodic Boundary Discontinuity (PBD)}

To quantify prosodic continuity at chunk boundaries, we extract F0 and energy in $\pm 1000 ms$ regions around each sentence boundary and compute: (1) $\Delta$ F0 (Hz jump) - absolute difference in mean F0 (Hz) before and after sentence boundaries, and (2) $\Delta$ Energy (dB) - energy discontinuity before and after sentence boundaries. We aggregate across all boundaries per passage and min-max normalize each of the aggregated metrics.

\textbf{Results:} From Table~\ref{tab:pbd} we find that our method achieves the best prosodic continuity, with boundary energy discontinuity of just 14.04 dB—roughly half that of competing models. While F0 jumps remain comparable across systems (67–69 Hz), energy consistency proves the dominant differentiator. XTTS, despite smooth pitch transitions, suffers from severe loudness inconsistency due to independent per-chunk gain normalization. Qwen3-TTS occupy the best middle ground with smooth transition in both pitch and energy. Our method's advantage stems from stateful inference maintaining coherence across chunks, which is the most perceptually salient aspect of boundary artifacts.

\begin{table}[t]
    \centering
    \caption{Prosodic Boundary Deviation (PBD) metrics across models. Lower values indicate better cross-sentence prosodic coherence. Composite is calculated by taking an average of $\Delta$ F0 and $\Delta$ Energy.}
    \label{tab:pbd}
    \resizebox{\columnwidth}{!}{
    \begin{tabular}{lccc}
        \toprule
        Model & $\Delta$ F0 (Hz) $\downarrow$ & $\Delta$ Energy (dB) $\downarrow$ & Composite $\downarrow$ \\
        \midrule
        \textbf{MagpieTTS-LF}  & 69.19 & \textbf{14.04} & \textbf{0.4646} \\
        XTTS~\cite{casanova2024xtts}  & 67.13 & 30.62 & 0.734 \\
        Qwen3-TTS~\cite{hu2026qwen3tts}  & \textbf{65.54} & 17.91 & 0.5169 \\
        VibeVoice~\cite{peng2025vibevoice}  & 69.08 & 28.90 & 0.712 \\
        \bottomrule
        \multicolumn{4}{l}{\scriptsize} \\
    \end{tabular}
    }
\end{table}

\begin{figure*}
    \centering
    \includegraphics[width=\textwidth, height=0.3\textheight, keepaspectratio=false]{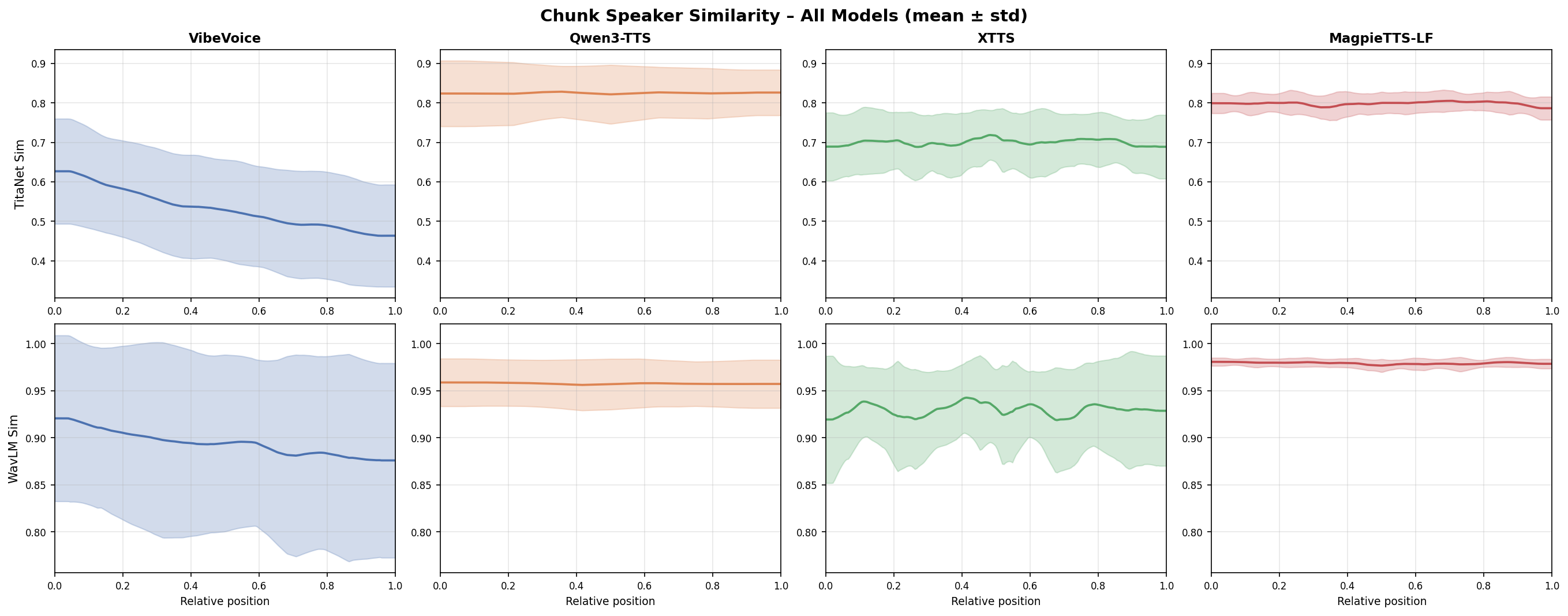}
    \caption{Speaker similarity (TitaNet, top; WavLM, bottom) across relative position in long-form utterances. Shaded regions denote standard deviation. MagpieTTS-LF maintains the most stable similarity throughout generation, while other models exhibit higher variance and drift over sequence length.}
    \label{fig:ssim}
\end{figure*}

\begin{figure*}[h!]
    \centering
    \includegraphics[width=1\linewidth]{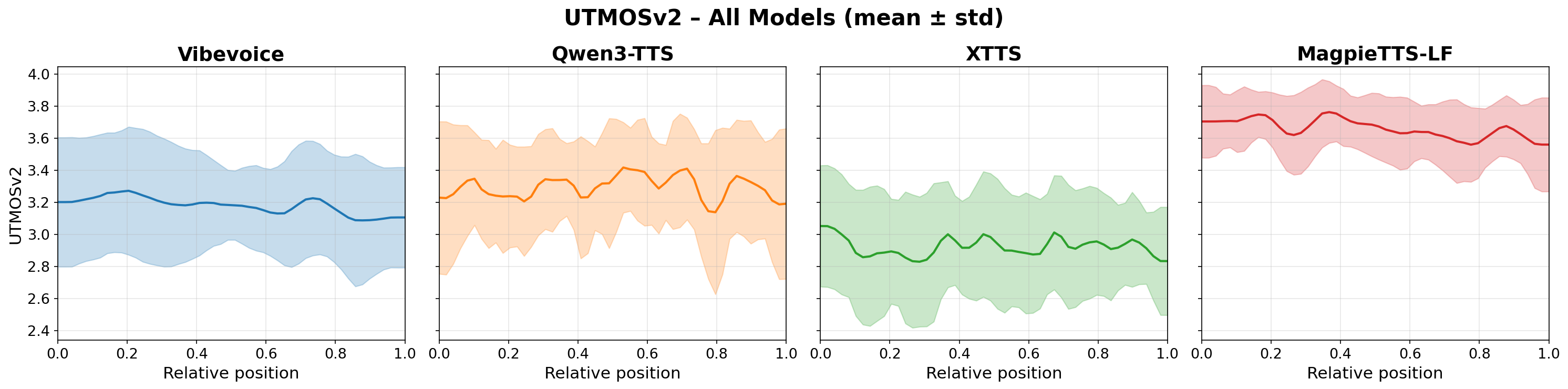}
    \caption{Plot of UTMOSv2 scores vs relative position in long-form utterances. Shaded regions denote standard deviation. MagpieTTS-LF achieves the highest quality with consistent scores throughout generation out-performing other baselines in UTMOSv2 score.}
    \label{fig:mos}
\end{figure*}

\subsection{Naturalness and Speaker Consistency}

To measure speaker consistency, we first divide the generated audio into non-overlapping chunks of 10 seconds each. We extract Titanet \cite{koluguri2022titanet} and WavLM embeddings \cite{Chen_2022} from each of the chunks and compute cosine similarity with reference speaker audio.  We plot the speaker similarity of these chunks against their relative positions in the long text. For example, a 10 second chunk close to the starting of the speech has a relative position closer to 0 on the x-axis and one closer to the end of the speech is closer to 1. To measure naturalness consistency we measure UTMOSv2 \cite{saeki2022utmos} from 10 second windows and visualize the change in UTMOSv2 score relative to position in longer sequences.

\textbf{Results:} As shown in Figure~\ref{fig:ssim}, MagpieTTS-LF maintains the highest and most stable speaker similarity throughout the sequence for both TitaNet and WavLM similarity. It also shows minimal variance and no observable drift from start to end, indicating that speaker characteristics are preserved across chunk boundaries. All competing models exhibit either higher variance, or drift showing speaker characteristics were not preserved across boundaries and over long distances. Interestingly, VibeVoice shows a downward trend, indicating inconsistent speaker representation despite single-pass generation.
Magpietts-LF has the highest UTMOSv2 and lowest variance, which shows that MagpieTTS-LF produces the most natural and stable audio. VibeVoice degrades the most over the course of long-form synthesis, XTTS produces the least natural sounding speech.

\section{Conclusion}

We present MagpieTTS-LF, an inference-time approach to synthesize robust, coherent and natural sounding long-form speech without retraining on long-form data. Our method uses soft attention prior to guide monotonicity, a history-aware stateful chunk generation that helps maintain prosodic continuity and speaker consistency over the entirety of the generated speech.
We also demonstrate that MagpieTTS-LF achieves much lower word error rates than competing systems, the lowest prosodic boundary discontinuity, and stable speaker similarity throughout generation compared to the state-of-the-art long-form TTS systems. MagpieTTS-LF also maintains the highest and most stable naturalness. We show these by experimenting with multiple dimensions of generated speech. By operating entirely at inference time, our approach can be extended to any chunk-based encoder-decoder TTS system for long-form synthesis.

\section{Generative AI Use Disclosure}

Generative AI was used for checking grammar and spelling of the entire paper. It was very minimally used to refine the language at some parts of the paper and with \LaTeX syntax.

\bibliographystyle{IEEEtran}
\bibliography{mybib}

\end{document}